# To accept or not to accept? An IRT-TOE Framework to Understand Educators' Resistance to Generative AI in Higher Education

Jan-Erik Kalmus and Anastasija Nikiforova

**Abstract:** Since the public release of Chat Generative Pre-Trained Transformer (ChatGPT), extensive discourse has emerged concerning the potential advantages and challenges of integrating Generative Artificial Intelligence (GenAI) into education. In the realm of information systems, research on technology adoption is crucial for understanding the diverse factors influencing the uptake of specific technologies. Theoretical frameworks, refined and validated over decades, serve as guiding tools to elucidate the individual and organizational dynamics, obstacles, and perceptions surrounding technology adoption. However, while several models have been proposed, they often prioritize elucidating the factors that facilitate acceptance over those that impede it, typically focusing on the student perspective and leaving a gap in empirical evidence regarding educators' viewpoints. Given the pivotal role educators play in higher education, this study aims to develop a theoretical model to empirically predict the barriers preventing educators from adopting GenAI in their classrooms. Acknowledging the lack of theoretical models tailored to identifying such barriers, our approach is grounded in the Innovation Resistance Theory (IRT) framework and augmented with constructs from the Technology-Organization-Environment (TOE) framework. This model is transformed into a measurement instrument employing a quantitative approach, complemented by a qualitative approach to enrich the analysis and uncover concerns related to GenAI adoption in the higher education domain.

## 1 Introduction

Generative Artificial Intelligence (GenAI), a type of artificial intelligence (AI) systems, excels at generating diverse content, thereby offering unprecedented capabilities, however, raising ethical concerns regarding disinformation, accuracy, bias, privacy, and ownership [7, 9, 12, 6]. Academic institutions responded swiftly due to concerns about misuse, particularly academic fraud. Some governments restricted or even banned ChatGPT [35, 18], while others provided guidelines for its ethical use [1]. Digital technology in education, including AI systems, addresses challenges such as those posed by the COVID-19 pandemic and aims to enhance learning experiences [14], potentially contributing to the Education 5.0 movement. However, integrating such technology requires understanding both its benefits and challenges. Research typically emphasizes identifying factors facilitating acceptance rather than barriers hindering it [28, 29, 15, 11, 30], crucial for promoting responsible adoption. This focus often centers on the student perspective, neglecting empirical evidence on educators' views, who are key actors in the educational ecosystem. Furthermore, research tends to concentrate solely on ChatGPT, disregarding other GenAI tools (see Section 2).

This study aims to develop a theoretical model to empirically predict barriers hindering educators from adopting GenAI in their classrooms. Considering a lack of models focusing on these barriers, the developed model integrates Innovation Resistance Theory (IRT) with elements of the Technology-Organization-Environment (TOE) framework.

The remainder of the paper is structured as follows: Section 2 provides background, Section 3 presents the model and methodology, while Section 4 concludes the paper.

## 2  Background

To study technology adoption, researchers widely use various models and theories to understand factors influencing an individual's decision to adopt technology. In this context, *technology acceptance* refers to perceived usefulness and attitudes towards a technology, considered a preliminary step to *technology adoption*, defined as actual technology use [13].

Research in this field has produced many empirically validated models over decades. Some theories focus on the perception-based pre-adoption phase, primarily using the Theory of Planned Behavior (TPB), which argues that perceptions drive actions [26]. However, Schwartz et al. [26] criticize the field for overlooking post-adoption perspectives lack of examination on why technology adoption is resisted. These gaps hinder a holistic understanding of motives towards certain technologies, particularly reluctance towards their adoption.

Prominent theories used in technology acceptance research include Technology Acceptance Model (TAM), Unified Theory of Acceptance and Use of Technology (UTAUT), Theory of Planned Behavior (TPB), and Technology-Organization-Environment (TOE).

**TAM** posits that user attitude towards a system is the most accurate predictor of actual use, which is based on perceived usefulness and perceived ease of use, emphasizing user perception [10].

**UTAUT** [32] claims the likelihood of adopting a technology depends on performance expectancy, effort expectancy, social influence, and facilitating conditions. UTAUT2 added hedonic motivation, price value, and habit to the model [33].

**TPB** predicts and explains human behavior through perception, attitude, and control over behavior [2]. It improves upon the Theory of Reasoned Action (TRA), focusing on intention driven by attitude, subjective norm, and perceived behavioral control.

**TOE** framework posits that technology adoption in organizations is influenced by the characteristics of the technology, the organizational context, and the external environment [21].

While these models are valuable for understanding factors that facilitate technology acceptance, they do not primarily focus on identifying barriers to adoption. Understanding both acceptance and resistance aspects is essential for a comprehensive view of technology integration. Innovation Resistance Theory (IRT) plays a crucial role in understanding barriers and resistance factors overlooked by TAM, UTAUT, TOE, and TPB.

**IRT** aims to explain consumer response to innovation and why some innovations are met with resistance [22]. According to Ram, resistance is a natural response to change and products that respond to the sources of resistance are more likely to overcome the resistance barriers [22]. IRT accounts for *functional* and *psychological barriers*, where functional barriers are associated with value, patterns and risks associated with product use, while psychological barriers are related to the traditions, norms, and perceptions of the customer [23]. According to IRT, three major groups of factors affect innovation resistance: (1) *perceived innovation characteristics*, (2) *consumer characteristics* and (3) *characteristics of propagation mechanisms*. The theory has its roots in business and marketing research and contrary to the previous models and frameworks introduced in this section, IRT attempts to understand *why consumers or organizations reject technology instead of embracing it*.

Transitioning to the current state of research on GenAI adoption, our investigation delved into existing studies through a systematic literature review (SLR). While we do not present the SLR findings in this paper due to its peripheral role, it is accessible on Zenodo - 10.5281/zenodo.13122996. The analysis revealed that the majority of existing research predominantly focuses on ChatGPT as the GenAI tool of choice [28, 29, 15, 11, 30, 3] (as of December 2023), which is unsurprising. However, it also underscores a research gap, indicating the necessity to encompass a broader range of GenAI tools.

Most studies focus on students as their primary group [28, 29, 15, 11, 30], with only one study focusing on academics [3]. The most popular technology adoption model in these studies was UTAUT, with two studies using the updated UTAUT2 [28, 29], and one using the original version [15]. Additionally, two studies employed TAM [11, 30], and one used Social Cognitive Theory (SCT) as its theoretical foundation [3]. Existing studies have primarily examined factors affecting the acceptance of GenAI technology, neglecting research on resistance, barriers, and limitations to adoption. This aligns with [25], emphasizing the need to understand why technology adoption is resisted—a perspective largely overlooked.

Research on the adoption of GenAI tools by academics and educators is notably limited. Doung et al. [11] have called for such research, which is particularly relevant in settings where educators have the final say on the use of these technologies (e.g., Estonian higher education). Given the identified research gaps, this study aims to explore GenAI adoption through the lens of resistance and educator perception, thereby being unique in all three components.

## 3 Model development

The study aims to uncover barriers hindering academics and higher education staff from adopting GenAI tools in classrooms. To this end, we develop a model integrating Innovation Resistance Theory (IRT) with Technology-Organization-Environment (TOE) constructs, considering both individual sentiment and organizational context. The model, rooted in IRT, is expanded with *Organization* and *Environment* constructs from TOE to incorporate institutional and external factors. The *Technology* construct is omitted to prevent redundancy as IRT captures the potential limitations of the technology from the perspective of the user.

Let us introduce the model's constructs (Fig. 1), where hypotheses and measurement items (Table 1) are formulated based on the underlying constructs and theory, existing research, and assumptions drawn from the current state of the art, providing a structured approach to hypothesis development.

### 3.1 Hypotheses development

The initial reception of GenAI by the education community can be perceived negative as implies from several bans and concerns of cheating conveyed through mainstream media [7, 9, 5, 35]. This leads us to **H1**: *More than half of educators prohibit the use of GenAI tools in their courses*.

The first of the psychological barriers in IRT – the **tradition barrier (TB)** – refers to the psychological impact of an innovation's deviation from the established traditions or societal norms of its user [23]. The intent to maintain existing processes and norms affects openness to change as change presents a threat to the status quo. The tradition barrier signifies resistance to change stemming from entrenched norms and practices, which Shills defines as "*anything which is handed down from past to present*" [27, p. 112]. Academic institutions, proud of their traditions and heritage, are often resistant to change, driven by concerns over disrupting established processes, workload implications, and distrust towards administration [4]. This reluctance is compounded by the autonomy of teaching staff. Organisation for Economic Cooperation and Development (OECD) echoes these concerns, suggesting that they may challenge traditional teaching methods and alter perceptions of education, impacting student engagement [20]. Consequently, we define **H2**: *Tradition barrier has a positive effect on academics' resistance towards the adoption of GenAI within the academic environment*.

The **image barrier (IB)** addresses how users perceive an innovation's impact on their social image or identity [23]. Bin-Nashwan et al. [3] exemplify this through academics' negative sentiment towards ChatGPT, driven by their dedication to academic integrity, which is integral to their personal reputation and self-identity. Consequently, we define **H3**: *Image barrier has a positive effect on academics' resistance towards the adoption of GenAI within the academic environment*.

The **risk barrier (RB)** encompasses the emotions stemming from uncertainty, unintended outcomes, or side-effects of innovation, categorized into physical, economic, functional, and social risks [23]. In the context of GenAI in higher education, concerns about academic fraud and cheating are prominent. Kasneci et al. highlight various issues, including overreliance, adaptability, and output verification, associated with GenAI use in education [16]. Additionally, GenAI tools are perceived as susceptible to producing inaccurate information and bias [16, 24]. Considering these risks, academics may resist GenAI adoption within the academic environment. For instance, the potential inaccuracy of credit point representations due to significant assistance from GenAI tools could impact the perceived value of students' efforts, rendering current estimations inaccurate. This leads us to **H4**: *Risk barrier has a positive effect on academics' resistance towards the adoption of GenAI within the academic environment*.

The **usage barrier (UB)** represents resistance to innovation when it clashes with existing workflows, practices, and habits [23]. In the context of GenAI in education, one significant issue is the challenge of discerning GenAI-generated outputs from student-produced work [16]. This poses difficulties in evaluating students' knowledge against course learning outcomes, as exemplified by the deprecation of essays as an evaluation method due to students' ability to generate them easily with GenAI tools. Addressing this challenge requires educators to adapt and improve evaluation methods [19] as done by a number of top-ranked universities worldwide, which may demand significant effort depending on course content and structure. Moreover, the usage barrier encompasses the incompatibility of GenAI within specific courses, such as language courses focused on reading and writing, where GenAI capabilities may exceed course requirements. Thus, **H5** *Usage barrier has a positive effect on academics' resistance towards the adoption of GenAI within the academic environment.*

The **value barrier (VB)** asserts that customers are unlikely to embrace a product if its performance does not justify its cost [23]. In the educational context, educators lack control over financial decisions affecting students' access to GenAI tools. Therefore, this study considers the value barrier from the perspective of the intrinsic value of the technology in education—whether its use in courses provides any value to students at all.

Effective technology integration in education necessitates equal access for all students, ensuring digital equity in classrooms [8, 12]. Historically, concerns about access and equity have been raised regarding the widespread availability of the internet in education [17], similar to the adoption of GenAI. Educators recognize the benefits of technology but emphasize the importance of equal availability and access. Within this barrier, it's crucial to weigh the required effort against the technology's potential benefits in a given course. Considering these factors allows for a more holistic understanding of the value barrier within this study's context. This leads us to **H6**: *Value barrier has a positive effect on academics' resistance towards the adoption of GenAI within the academic environment.*

The **organizational** context (OB) within the TOE framework considers how an organization's staff, internal structure, and processes influence the adoption of technology [31]. In this study, organization refers to the educational institutions where educators are employed. Webb and Cox identified that for effective technology integration in classrooms, teachers require knowledge on how to effectively incorporate these technologies into teaching practices [34]. Institutions may need to provide additional support to educators in the form of knowledge sharing, training, and guidelines to facilitate these changes. Additionally, educators may feel pressured to adopt these tools due to their increasing popularity and promotion as the future of education, which could lead to resistance. This leads us to **H7** - *Organisational context has a positive effect on academics' resistance towards the adoption of GenAI within the academic environment.*

The **environment** context (EB) within the TOE framework considers the impact of external variables such as regulation, market conditions, and societal and cultural factors [31]. Various governmental restrictions have been observed on the use of ChatGPT since their public availability. For example, Italy initially prohibited the use of ChatGPT over privacy concerns [18], and local governments in Australia and the United States also imposed restrictions on its use in education [5, 35]. These examples highlight the susceptibility of GenAI technology to government regulation. Additionally, educators may feel pressured to integrate GenAI tools into their courses due to the technology's popularity and its increasing use in other fields, which could lead to resistance. This leads us to **H8**: *Environment context has a positive effect on academics' resistance towards the adoption of GenAI within the academic environment.*

### 3.2 Model

The developed model (see Figure 1) uses the constructs of the IRT and the *Environment* and *Organisation* constructs of the TOE models to capture the potential resistance sentiment of educators towards GenAI tools use in the classroom. The model uses demographic variables – gender, age, university, academic field – as control variables.

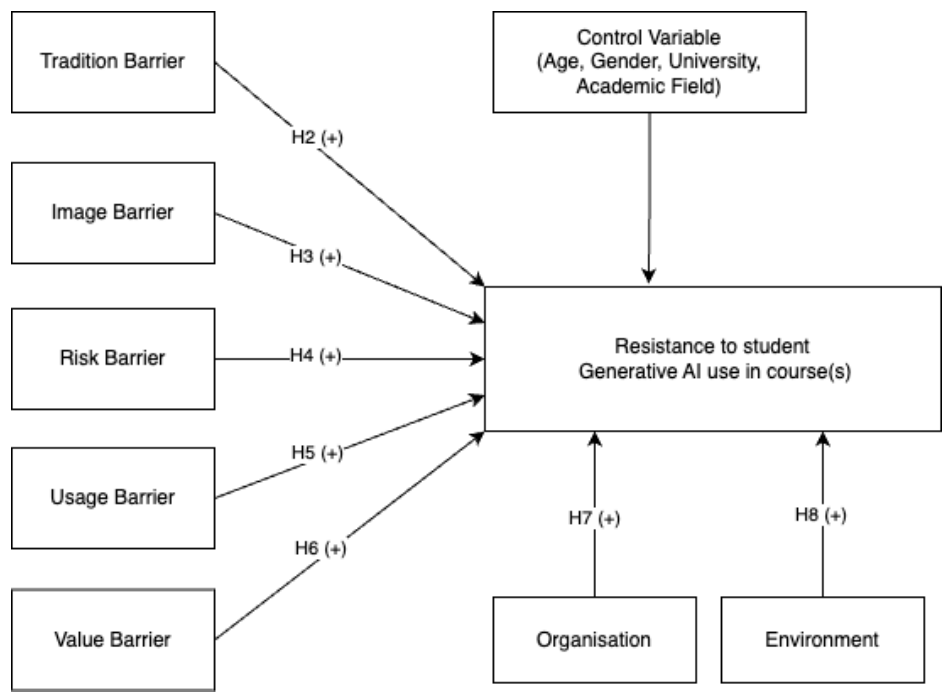

**Figure 1.** Consolidated theoretical model.

We propose a mixed-method approach to validate the hypotheses and refine the developed theoretical model. Utilizing Structural Equation Modelling (SEM) and statistical analysis, the quantitative method will test the accuracy of the model and identify potential areas for improvement. To gain deeper insights, we recommend complementing this with qualitative data collection, which is expected to be analyzed using response coding technique.

To this end, we develop a questionnaire for data collection (available on Zenodo - 10.5281/zenodo.13122996), which also gathers qualitative insights from educators on their perspectives regarding GenAI use. The survey is divided into three sections: (1) demographic profile, (2) core measurement instrument, and (3) additional questions. This mixed-method approach ensures a robust analysis and a comprehensive understanding of the factors influencing GenAI adoption in education.

The demographic profile section focuses on collecting demographic data related to gender, age range, university, and general academic field. In addition to the initial closed-ended questions aimed at recording the respondents' demographic profile, two additional closed-ended questions are included in the general information section: *Do you allow students to use GenAI in your courses? Do you use GenAI in your academic work?*

The first question is used to capture data for the H1. To reflect the sentiment towards using GenAI tools in courses more accurately, the possible answers are: *No*, *Allow*, and *Recommend*. The second question gathers information on the number of educators using GenAI in their academic work, helping to distinguish between respondents with prior experience using the tools and those whose perceptions may be influenced by a lack of familiarity with GenAI.

The second part of the questionnaire contains the measurement instrument consisting of 31 measurement items (see Table 1) derived from the constructs they are associated with based on existing research and underlying theory, wherein measurement items are measured using a five-point Likert scale.

**Table 1.** Developed measurement instrument

| |
|---|
| **TB1**: Student use of GenAI reduces student participation in traditional course delivery methods |
| **TB2**: Student use of GenAI makes traditional teaching methods ineffective |
| **TB3**: Student use of GenAI creates a need for new teaching methods |
| **TB4**: Student use of GenAI conflicts with general academic norms or traditions of my institution |
| **IB1**: Allowing students use GenAI in my courses causes criticism from my colleagues |
| **IB2**: Allowing students use GenAI in my courses has a negative effect on my academic reputation |
| **IB3**: Not allowing students use GenAI in my courses has a negative effect on my reputation among students |
| **RB1**: Students using GenAI can become overly reliant on the tools |
| **RB2**: Student use of GenAI has a negative effect on the development of problem solving and critical thinking skills |
| **RB3**: Students using GenAI in my courses can complete the course with reduced effort |
| **RB4**: Students using GenAI risk their academic integrity |
| **RB5**: Students using GenAI can receive misleading information from the tools |
| **RB6**: Students may have difficulties verifying the accuracy of GenAI outputs |
| **UB1**: The contents of my courses are not suitable for GenAI use by students |
| **UB2**: Student produced work is difficult to distinguish from GenAI outputs |
| **UB3**: Students use of GenAI makes measuring learning outcomes in my courses more difficult |
| **UB4**: Current evaluation methods in my courses are less effective when GenAI is used by students |
| **UB5**: Improved evaluation methods in my courses are necessary when GenAI is used by students |
| **VB1**: Students do not have equal access to GenAI tools that are useful in my courses |
| **VB2**: Students do not receive significant value from the use of GenAI tools in my courses |
| **VB3**: The risk of misuse of GenAI tools outweighs its potential benefits |
| **VB4**: The resources required to allow the use of GenAI tools outweigh its potential benefits |
| **O1**: Institutional policy is required to allow student GenAI use |
| **O2**: Current institutional policy does not support student GenAI use in courses |
| **O3**: My institution does not promote student GenAI use |
| **O4**: My institution does not provide me support (e.g., guidance, training) to allow students GenAI use in my courses |
| **O5**: My institution does not provide students support (e.g., guidance, training) to allow them use GenAI in my courses |
| **E1**: Government policy is required to allow the use of GenAI tools by students |
| **E2**: Current government policy does not support student GenAI use |
| **E3**: My institution expects me to allow students use GenAI tools in my courses |
| **E4**: Students expect me to allow them use GenAI tools in my courses |

The questionnaire is complemented with open-ended questions to gain a descriptive insight into the general sentiment of educators towards GenAI, covering multiple aspects of educators' personal experience with GenAI from personal use to necessary adjustments in courses.

The third section includes optional open-ended questions to provide respondents with a platform for a more nuanced interpretation of results. These questions inquire:

- *Are there any additional challenges associated with letting students use GenAI in your course(s) that were not covered by this study?*
- *What kind of tasks in your courses are most affected by the use of GenAI?*
- *How would you describe your own experience of using GenAI tools?*
- *Have you already integrated the use of GenAI in your course(s)?*
- *What were the adaptations you already did to integrate GenAI use in your course(s)?*

The instrument was subject to iterative expert review throughout its development process to improve the quality of the measurement items. Reviews were performed by two experts from the domains of technology acceptance research and modern educational theory. The instrument is made available on Zenodo - 10.5281/zenodo.13122996.

## 4 Conclusion

This study proposes a novel theoretical model aimed to uncover barriers preventing academic staff from adopting GenAI tools in their courses. Drawing on constructs from Innovation Resistance Theory (IRT) and Technology-Organization-Environment (TOE), it examines individual, organizational, and environmental aspects to empirically assess potential resistance to GenAI use by students. Utilizing these constructs and the state of the art, we define eight hypotheses regarding resistance to GenAI use in higher education. The model is transformed into a measurement instrument that employs quantitative and qualitative approaches to enrich the analysis and discover concerns related to GenAI adoption in higher education.

Future research is recommended to validate the developed model in real-world settings with its further refinement. As such, we are currently applying the model in Estonian higher education settings to examine whether academic staff in Estonia, known as a "digital nation" and early adopters of emerging technologies like GenAI, are reluctant to allow students to use GenAI tools, identifying concerns related to GenAI adoption in higher education.

The study contributes to the research by proposing a novel theoretical model that addresses the often-neglected perspective of technology resistance. It is one of the first to encompass educators' viewpoints as integral to the educational ecosystem. However, while we cover a neglected perspective, further research would benefit from unifying student and educator perspectives to comprehensively understand factors affecting GenAI adoption in higher education.

### Declaration of Generative AI use

The authors hereby disclose that Google Gemini was used to improve the conciseness and clarity of selected sentences in this paper.